\documentclass[a4paper,
               ]{jacow}
\usepackage{caption}
\usepackage{subcaption}
\usepackage{pdfpages,multirow,ragged2e} %
\usepackage{fancyhdr}  
\usepackage[utf8]{inputenc} 
\usepackage{calc}
\usepackage[mathlines, switch]{lineno}

\makeatletter%
	\ifboolexpr{bool{xetex}}
	 {\renewcommand{\Gin@extensions}{.pdf,%
	                    .png,.jpg,.bmp,.pict,.tif,.psd,.mac,.sga,.tga,.gif,%
	                    .eps,.ps,%
	                    }}{}
\makeatother

\ifboolexpr{bool{xetex} or bool{luatex}} 
 {}                                      
 {\usepackage[utf8]{inputenc}}           

\usepackage[USenglish]{babel}

\begin{document}

\title{
  \vspace{-19pt}
  \small Presented at the 32nd International Symposium on Lepton Photon Interactions at High Energies, Madison, Wisconsin, USA, August 25-29, 2025. \\
  
 \Large Measurement of multi-jets and vector boson plus jets production in ATLAS}

\author{Giulia Manco\thanks{gmanco@cern.ch. \\Copyright 2025 CERN for the benefit of the ATLAS Collaboration.
CC-BY-4.0 license.}, on behalf of the ATLAS Collaboration. \\ \textit{Pavia University and INFN,
5 Via Agostino Bassi, 6, 27100 Pavia PV, Italy.} \\
}
	
\maketitle

\begin{abstract}
   \noindent The production of multiple jets or vector bosons in association with jets at the LHC provides a unique testing ground for Quantum Chromodynamics (QCD) in the high-energy regime. 
   With the increasing precision of the ATLAS measurements, detailed studies have become possible on observables that probe different aspects of QCD, such as the topological configurations between vector bosons and jets, jet substructure features, and heavy-flavor jet contributions. 
   These measurements also play a key role in improving the precision of the determination of the strong coupling constant.
   Recent ATLAS results in these areas are presented, offering new insights into QCD dynamics and the performance of state-of-the-art theoretical predictions.
\end{abstract}

\section{Introduction}

\noindent 
The production of multiple jets or vector bosons in association with jets is fundamental for probing and improving our understanding of the strong interaction, which plays a central role in proton--proton (\textit{pp}) collisions at the LHC. 
Jets, originating from the fragmentation of high-energy quarks and gluons, provide access to QCD phenomena across different regimes, as they encompass both perturbative and non-perturbative contributions.
These measurements explore the strong interaction through jet substructure observables, which probe correlations among the particles produced within an identified jet. \\
At the ATLAS experiment~\cite{ATLAS_detector}, several analyses in recent years have investigated various jet substructure variables. \\
The measurement of the transverse momentum fraction, $r_q$, carried by charged hadrons resulting from the fragmentation of quarks or gluons, using dijet events at $\sqrt{s} = \SI{13}{\TeV}$~\cite{ATLAS:2923297}, provides direct information on track-based jet substructure, known as \textit{track functions}.
Another important jet substructure observable is the \textit{Lund Jet Plane}, which reconstructs the hierarchy of QCD emissions inside the jet and offers insights into parton shower dynamics and hadron formation. 
In ATLAS, the Lund Jet Plane has been measured for the first time in $t\bar{t}$ events, using the full Run~2 dataset of \SI{140}{fb^{-1}} of $\sqrt{s} = \SI{13}{\TeV}$ \textit{pp} collisions~\cite{ATLAS:LJP}.\\
Measurements of vector boson production in association with multiple jets are also sensitive to both QCD and electroweak (EW) corrections. \\
In particular, the measurement of collinear $W$-boson emission from high-$p_\mathrm{T}$ jets at $\sqrt{s} = \SI{13}{\TeV}$, using \SI{140}{fb^{-1}} of data~\cite{ATLAS:2920036}, provides a stringent test of higher-order theoretical predictions. 
Finally, the measurement of the production cross-section of a $Z$ boson in association with $b$- or $c$-jets in \textit{pp} collisions at $\sqrt{s} = \SI{13}{\TeV}$ with \SI{140}{fb^{-1}}~\cite{ATLAS:Z+c} offers an important test of perturbative QCD and of the internal structure of the proton.

\section{Measurement of Jet Track Functions}
\begin{figure}[!htb]
  \centering
 \begin{subfigure}[b]{0.4\textwidth}
   \centering
   \includegraphics[scale=0.2]{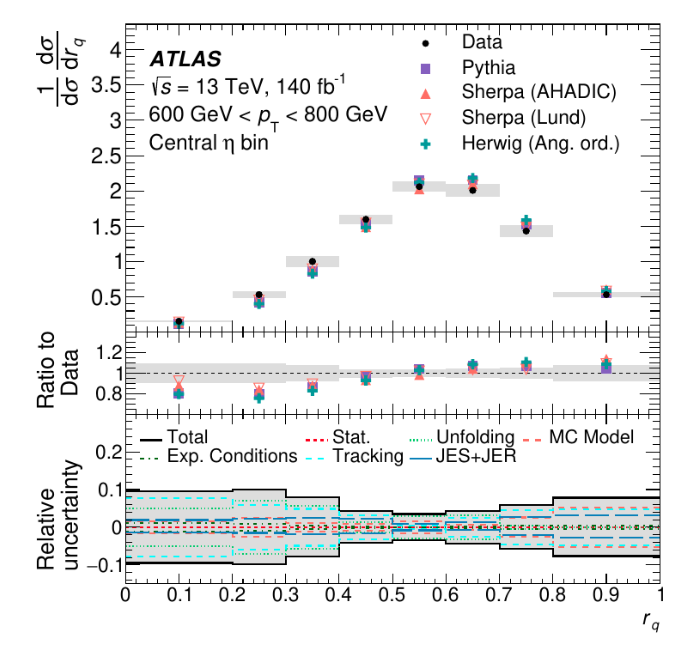}
   \caption{}
   \label{fig:rq_central}
  \end{subfigure}
\hfil 
  \begin{subfigure}[b]{0.4\textwidth}
    \centering
    \includegraphics[scale=0.2]{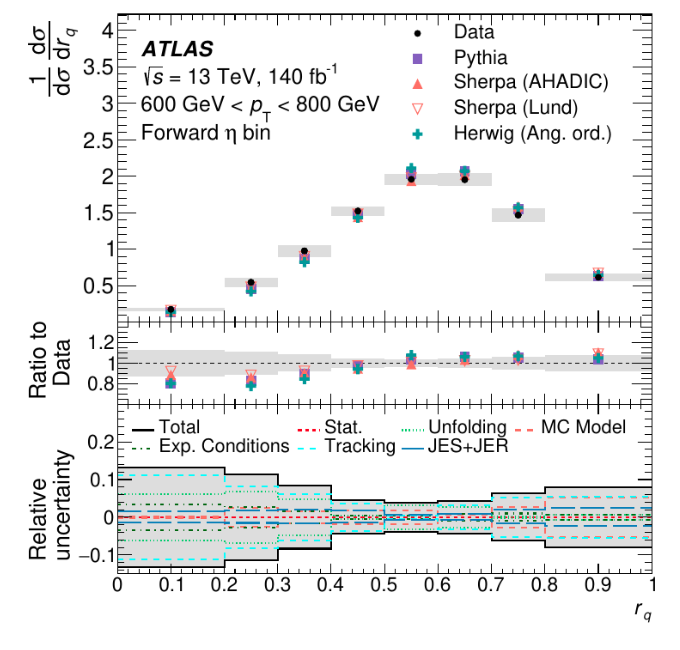}
    \caption{}
    \label{fig:rq_forward}
  \end{subfigure}
  \caption{The unfolded central (a) and forward (b) normalized differential cross-sections as a function of $r_q$ for data compared to predictions from several MC generators \cite{ATLAS:2923297}.}
  \label{fig:rq}
\end{figure}

\noindent 
The hadronization processes in jets, from a theoretical point of view, are characterized by the interplay between perturbative and non-perturbative functions, which are universal within the framework of factorization. 
The functions describing the fragmentation of quarks and gluons into charged hadrons, identified in the detector as tracks, are particularly important, since track-based measurements can provide more precise results thanks to the superior angular granularity of the tracking detectors. 
\\The correlated fragmentation of quarks and gluons into charged hadrons is described by universal non-perturbative functions known as \textit{jet track functions}. 
These observables characterize the energy distribution of charged hadrons and, unlike the well-known fragmentation functions, they are multi-hadron observables that do not exhibit a linear Renormalization Group (RG) scale evolution, i.e.\ they go beyond the DGLAP paradigm. 
\\Experimentally, the jet track functions can be determined from the measurement of the transverse momentum fraction of tracks in high-energy jets, defined as
\[
r_q = \frac{p_{T}^{\mathrm{charged}}}{p_{T}^{\mathrm{all}}},
\]
where $p_{T}^{\mathrm{all}}$ is the transverse momentum of the jet and $p_{T}^{\mathrm{charged}}$ is the transverse momentum of the sum of charged particles within the jet.
\\The measurement of the energy distribution of tracks in identified high-$p_T$ jets, using the full Run~2 dataset of proton--proton collisions at $\sqrt{s} = 13~\mathrm{TeV}$ recorded with the ATLAS detector, is performed differentially in $r_q$ and in different regions of jet $p_T$, as well as for different rapidity orderings of a dijet system (i.e., for central and forward jets)\cite{ATLAS:2923297}. 
The distributions are used to extract the moments of $r_q$ as a function of jet $p_T$. The data are corrected using an Iterative Bayesian Unfolding (IBU) method and a machine learning-based method, OmniFold, which is also employed for moment extraction. \\
The unfolding is implemented simultaneously in three dimensions (jet $p_T$, $\eta$, and $r_q$). 
In Figure~\ref{fig:rq}, the unfolded central(~\ref{fig:rq_central}) and forward(~\ref{fig:rq_forward}) normalized differential cross-sections as a function of $r_q$ are shown for data and compared to predictions from several MC generators, which differ in their hadronization models. The measured cross-section is shown for a representative bin of jet $p_T$ of 600--800~GeV. 
Overall, data and MC are in general agreement within the uncertainties; however, some MC generators tend to underestimate the cross-section at low values of $r_q$ and overestimate it at high values. 
\\The grey uncertainty band represents the combination of statistical and systematic uncertainties. It is evident that the tracking uncertainty is dominant. 
From the $r_q$ distributions, the moments are extracted as a function of jet $p_T$. The moments are found to be slowly dependent on jet $p_T$. 
The moments are calculated as the sum over the $r_q$ bin contents, which introduces a discretization effect. 
A data-driven unbinned method, implemented in OmniFold, allows a correction on the reconstructed sample using the truth sample as input for a neural network.
\\The non-linear RG evolution of the track functions can be investigated through the cumulant pairs $k_i$, which are functions of the moments. 
The energy dependence of the relationship between cumulants is theoretically determined by the RG flows, providing insight into correlations in the hadronization process. 
In this analysis, the pairs of cumulants extracted from the unfolded data are compared to an analytical QCD prediction at next-to-leading logarithm (NLL) accuracy of the RG flow. 
The results show good agreement with theory, though they are not yet fully conclusive and require further investigation.

\section{Lund Jet Plane in hadronic decays of top and W}
\begin{figure}[!htb]
   \centering
   \includegraphics*[width=.68\columnwidth]{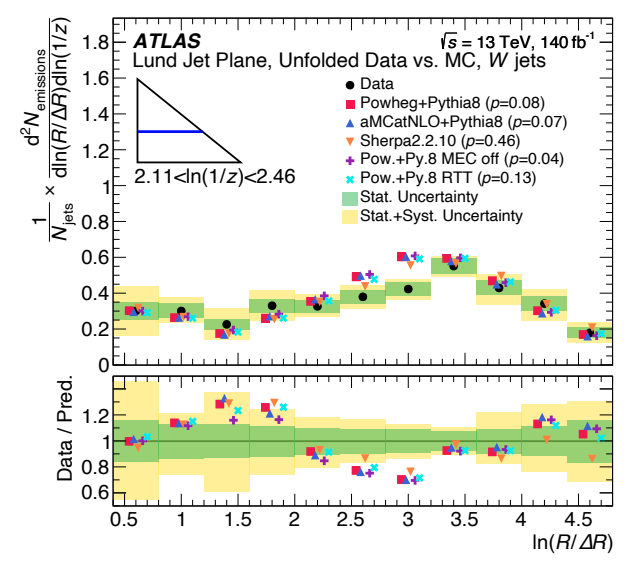}
   \caption{Slice of measured W jet LJP as function of $\ln(R/\Delta R)$ \cite{ATLAS:LJP}.}
   \label{fig:LJP}
\end{figure}
\noindent The Lund Jet Plane (LJP) observable provides valuable information about the process of jet formation, as it reconstructs the hierarchy of QCD emissions inside jets.\\
The first measurement of the LJP for jets initiated by top quarks and $W$ bosons has been performed by ATLAS, using the full Run~2 proton--proton dataset at a center-of-mass energy of $\sqrt{s} = 13~\mathrm{TeV}$ \cite{ATLAS:LJP}.\\
Since hadronic top-quark decays are challenging to measure and model due to their highly boosted decay products, jet substructure observables such as the LJP are particularly useful, as they are sensitive to different perturbative and non-perturbatives regimes.\\
The Lund Jet Plane is a two-dimensional representation of the jet substructure, showing the transverse momenta and angles of each emission with respect to the jet axis.\\
The LJP is constructed by reclustering emissions using the Cambridge--Aachen algorithm.\\
The LJP has recently been explored by several analyses in ATLAS, CMS, and ALICE, and plays an important role—alongside other jet substructure variables—in the development of jet-tagging algorithms.\\
The analysis presented here uses $t\bar{t}$ events, applying selections on the jet mass and the distance from a $b$-jet in order to categorize the jet as either top quark or $W$ boson.\\
One large-$R$ jet ($R = 1.0$) as $W$ boson or top quark with transverse momentum above 350~GeV is selected per event, together with one small-$R$ $b$-jet ($R = 0.4$) originating from the top-quark decay.\\
The parton shower and hadronization in these jet topologies are of particular interest due to the large masses of the top and bottom quarks, as well as the color-singlet nature of the $W$ boson.\\
The unfolded Lund Jet Plane distributions are compared with several MC generator samples, produced at NLO in QCD for the parton shower, or using alternative configurations of the nominal setup.\\
In Figure~\ref{fig:LJP}, the per-large-$R$-jet emission density as a function of $\ln(R/\Delta R)$ is shown for a single slice of $\ln(1/z)$, where $z$ is the transverse momentum fraction of the emission with respect to the jet, and $\Delta R$ is the emission angle.\\
The lower panel of the figure shows the data-to-prediction ratio, including both total and statistical uncertainties.\\
For each prediction, the compatibility with the data is estimated through a $\chi^2$ test, and the corresponding $p$-value is reported in parentheses.\\
For $W$ jets, no generator achieves full agreement with the measurement across the entire LJP phase space, although better agreement is observed in some subregions.\\
For top jets, the best description is obtained from the \texttt{Sherpa}~2.2.10 prediction.\\
The total uncertainty is dominated by $t\bar{t}$ modelling for most of the measured phase space.\\
This uncertainty is driven primarily by components related to the choice of hadronization model, parton shower, and the value of $\alpha^{\mathrm{FSR}}_s$.\\
These results provide valuable input for improving and tuning $t\bar{t}$ Monte Carlo event generators, particularly in parameters controlling radiation from the hard process, which are not yet well modelled in the LJP.\\

\section{Collinear $W$+jets}
\begin{figure}[!htb]
   \centering
   \includegraphics*[width=.9\columnwidth]{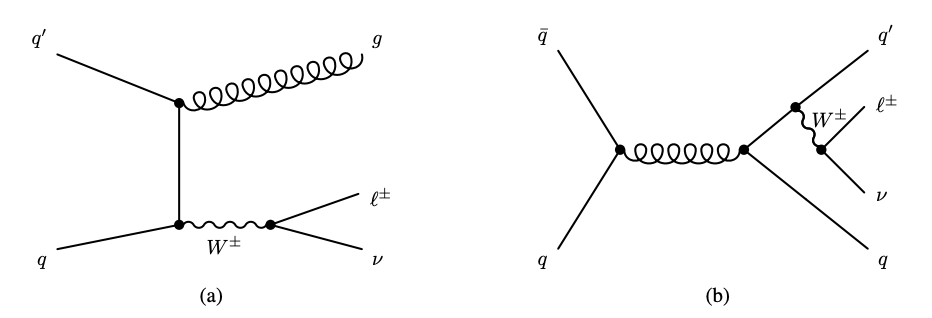}
   \caption{Feynman diagrams for $W$+jets production. In Figure (a), the $t$-channel shows the back-to-back region, while in Figure (b), the $s$-channel represents the collinear region.}
   \label{fig:Wjets}
\end{figure}
\noindent Measurements of $W$-boson production in association with high-momentum jets are presented using the full Run~2 proton-proton dataset at a center-of-mass energy of 13~TeV collected by ATLAS. \\
The measurement of the W+jets process aims to improve the precision of the $W$+jets process over a wide energy range. 
This process constitutes a significant irreducible background for several LHC analyses and provides a broad high-momentum phase space in which to study the modelling of kinematic variables. \\
In $W$+1~jet production at LO in $\alpha_s$, as shown in Figure~\ref{fig:Wjets}(a), the $W$ boson recoils against a quark or gluon, defining the so-called back-to-back region. 
The NLO in $\alpha_s$, as illustrated in Figure~\ref{fig:Wjets}(b), includes additional contributions, in particular, this analysis focuses on real $W$-boson emission from incoming or outgoing quarks accompanied by dijet emission. 
The latter contribution enhances the collinear region in the distribution of the angular separation between the $W$ boson and the closest high-$p_T$ jet. \\
The analysis measures the inclusive and differential cross sections of $W$ bosons decaying into electrons and muons produced in association with central jets with $p_T>30$~GeV, where the leading jet has $p_T>500$~GeV. 
This phase space is referred to as the \textit{inclusive region}.
The leptonic $W$-boson decay provides a clean experimental signature to test the EW sector and perturbative QCD in jet production. 
 \\
Several variables sensitive to collinear enhancement in the production rate are used for the differential measurements:
\begin{itemize}
  \item $\Delta R_{\mathrm{min}}(\ell, \mathrm{jet}^{100}_i)$ is the angular separation between the lepton from the $W$ boson and the closest jet with transverse momentum greater than 100~GeV. The region with $\Delta R_{\mathrm{min}}(\ell, \mathrm{jet}^{100}_i) >= 2.6$ is dominated by back-to-back events, while values below 2.6 correspond to the collinear region.
  \item $p_{T}^{\ell\nu}/p_{T}^{\mathrm{closest~jet}}$ is the ratio of the transverse momentum of the lepton–neutrino system to that of the jet closest to the lepton. The collinear region is defined by events where the transverse momenta are not similar, i.e.\ $p_{T}^{\ell\nu}/p_{T}^{\mathrm{closest~jet}} \neq 1$.
  \item $m_{jj}$ is the invariant mass of the two highest-momentum jets. This observable is often exploited in EW measurements and BSM searches and is challenging to model at high energies.
\end{itemize}
Differential measurements are also performed for other observables sensitive to $W$-boson production in association with multiple jets in the collinear region. \\
The unfolded data are compared to event generators accurate at NLO in $\alpha_s$, with and without EW corrections (\texttt{Sherpa}~2.2.11, \texttt{MadGraph5\_aMC@NLO+PYTHIA8}), merged with parton-shower models for higher-order emissions. Additionally, a fixed-order $W$+1~jet prediction at NNLO in $\alpha_s$ is performed using the \texttt{MCFM} program and compared with the data. \\

\begin{figure}[!hb]
   \centering
   \includegraphics*[width=.8\columnwidth]{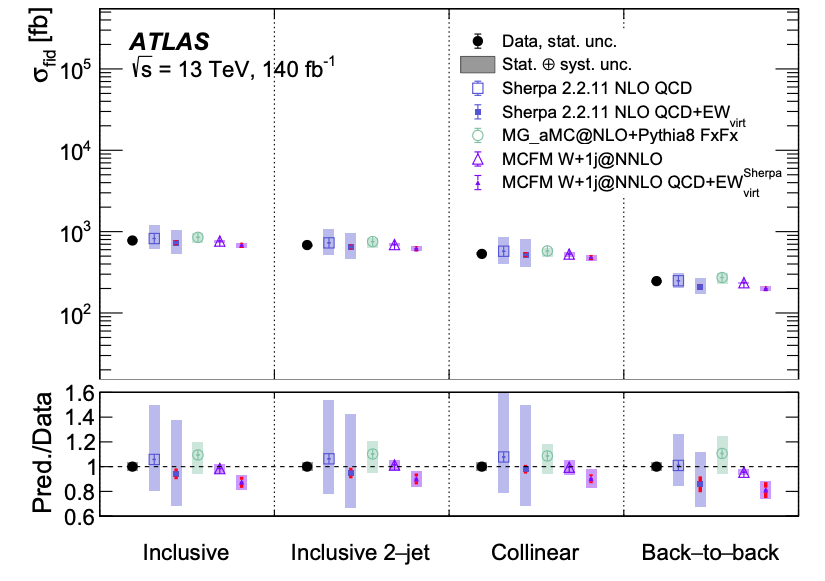}
   \caption{Measured W+jets fiducial cross-section in signal regions~\cite{ATLAS:2920036}.}
   \label{fig:Wjets_cx}
\end{figure}
\noindent Figure~\ref{fig:Wjets_cx} shows the total fiducial cross sections for different signal regions. The data correspond to a likelihood-fit combination of the electron and muon channels, taking into account uncertainty correlations.  
In addition to the \textit{inclusive region} already defined, the other signal regions are the \textit{inclusive 2-jets} region (requiring at least two jets), the \textit{collinear} region defined by $\Delta R_{\mathrm{min}}(\ell, \mathrm{jet}^{100}_i)<2.6$, and the \textit{back-to-back} region defined by $\Delta R_{\mathrm{min}}(\ell, \mathrm{jet}^{100}_i)>2.6$. \\
The data and their uncertainties are shown as solid dots with error bands; theoretical predictions are also displayed for comparison. 
\texttt{Sherpa} and \texttt{MadGraph} uncertainties include QCD scale and PDF variations, while the inclusion of EW corrections introduces additional uncertainties from different EW schemes. \texttt{MCFM} does not include PDF uncertainties. \\
In the lower panel, the prediction-to-data ratio is reported with the corresponding uncertainties.
\texttt{Sherpa} and \texttt{MadGraph} show larger uncertainties, mainly due to QCD scale variations. 
The \texttt{MCFM} prediction shows good agreement in all regions. 
The inclusion of NLO EW corrections improves the agreement of \texttt{Sherpa} in the collinear regions, while \texttt{MCFM} with NLO EW corrections slightly underestimates the data by about 10\%. \\
The total systematic uncertainty is around 3–4\%, dominated by uncertainties in background modelling and jet energy scale/resolution (JES/JER). \\
\begin{figure}[!htb]
  \centering
  \begin{subfigure}[b]{0.4\textwidth}
    \centering
    \includegraphics[scale=0.25]{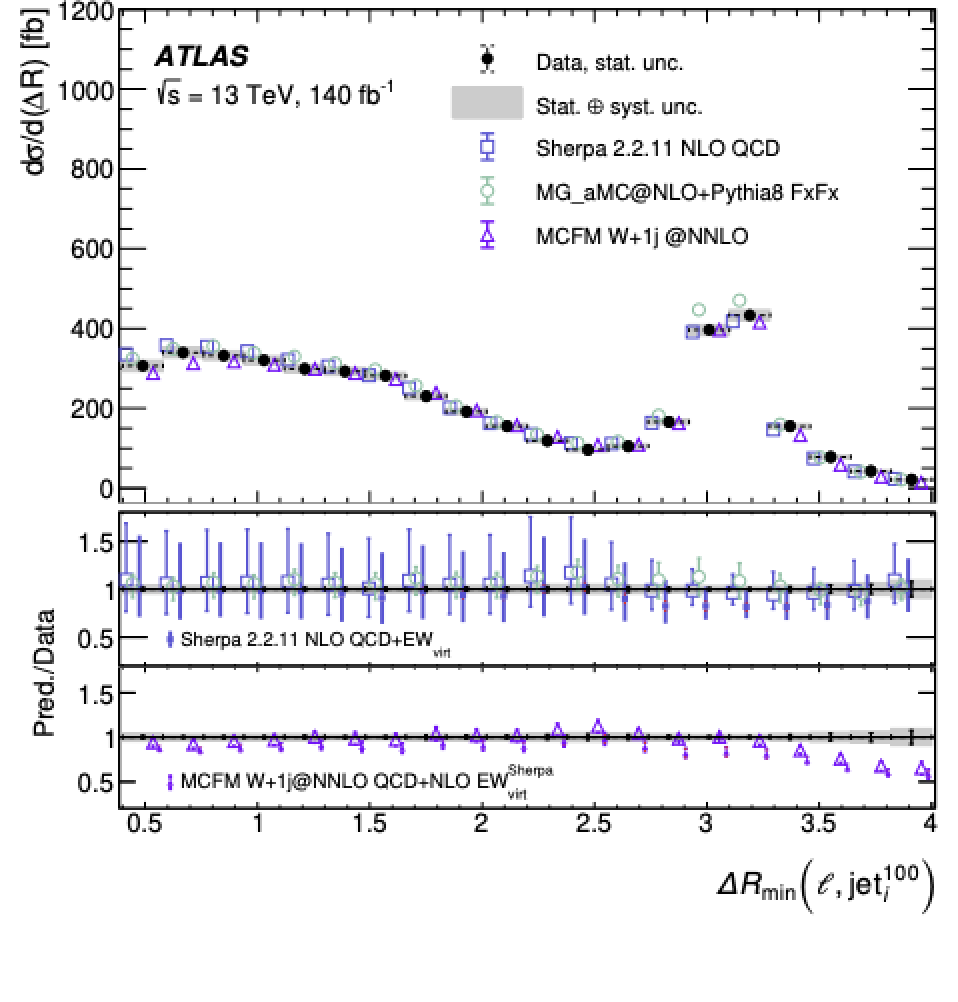}
    \caption{}
    \label{fig:WjetDR}
  \end{subfigure}
  \begin{subfigure}[b]{0.4\textwidth}
    \centering
    \includegraphics[scale=0.25]{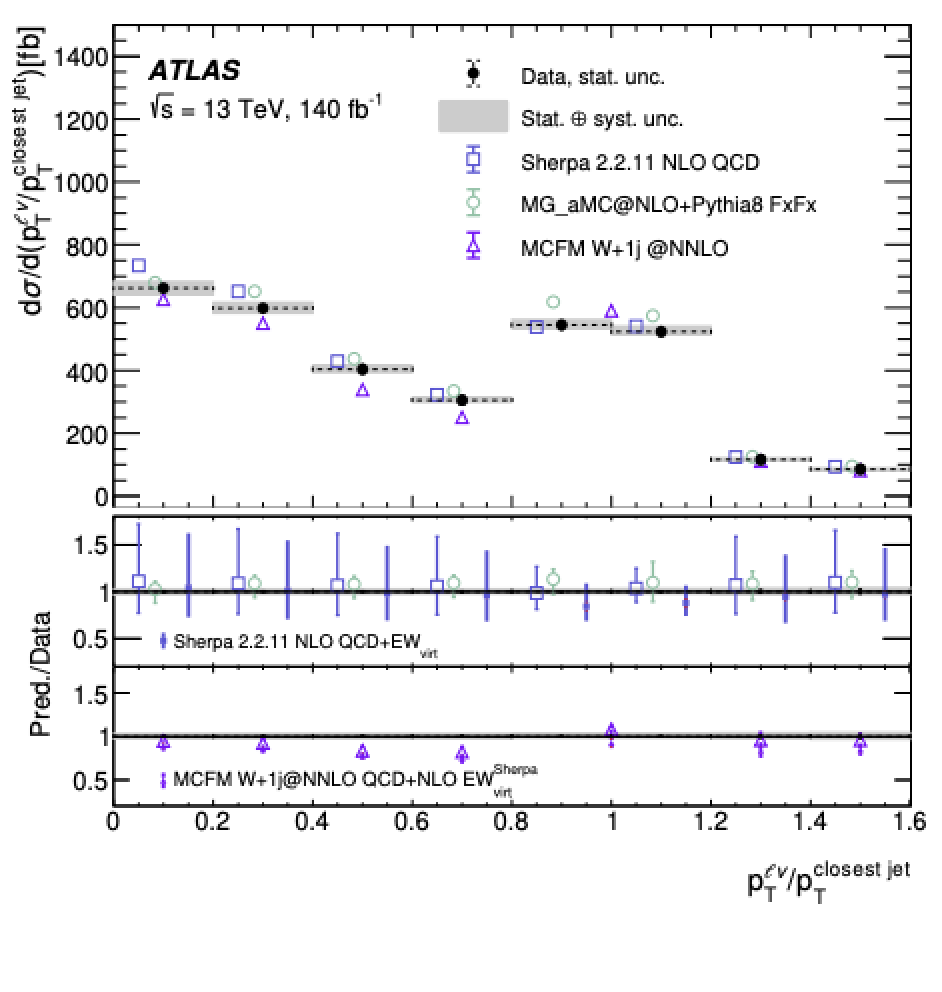}
    \caption{}
    \label{fig:WjetpT}
  \end{subfigure}
  \caption{Differential cross-section measurements in the inclusive phase space as a function of $\Delta R_{\mathrm{min}}(\ell, \mathrm{jet}^{100}_i)$ (a) and $p_{T}^{\ell\nu}/p_{T}^{\mathrm{closest~jet}}$ (b)~\cite{ATLAS:2920036}..}
  \label{fig:Wjets_diff}
\end{figure}

\noindent Figure~\ref{fig:Wjets_diff} shows the differential cross sections in the inclusive phase space as functions of $\Delta R_{\mathrm{min}}(\ell, \mathrm{jet}^{100}_i)$~(a) and $p_{T}^{\ell\nu}/p_{T}^{\mathrm{closest~jet}}$~(b). 
These variables help distinguish collinear events, where the $W$ boson is emitted by a high-momentum jet, from back-to-back events, where the $W$ boson recoils against the initial jet. 
The peak at $\Delta R_{\mathrm{min}}(\ell, \mathrm{jet}^{100}_i)=\pi$ arises from the leading-jet requirement of $p_T>500$~GeV, which, in back-to-back topologies, leads to the lepton being separated from the leading jet by approximately $\pi$. 
Such single-jet events contain a $W$ boson whose transverse momentum balances that of the leading jet, resulting in $p_{T}^{\ell\nu}/p_{T}^{\mathrm{closest~jet}} \approx 1$. \\
The data and their statistical uncertainties are shown as solid dots with error bars, while the combined statistical and systematic uncertainties are displayed as shaded bands. \\ 
Overall, \texttt{MadGraph5\_aMC@NLO+Pythia8 FxFx} provides a good description of both variables. 
Values of $\Delta R_{\mathrm{min}}(\ell, \mathrm{jet}^{100}_i)>\pi$ show some disagreement with \texttt{MCFM}. 
In the lower panel, the data are compared with the predictions including EW corrections. 
The inclusion of NLO EW corrections in \texttt{Sherpa} improves the agreement in the collinear region but underestimates the data in the back-to-back region. \\

\section{Measurement of $Z$+heavy-flavour jets}

\begin{figure}[!htb]
  \centering
  \begin{subfigure}[b]{0.2\textwidth}
    \centering
    \includegraphics[scale=0.18]{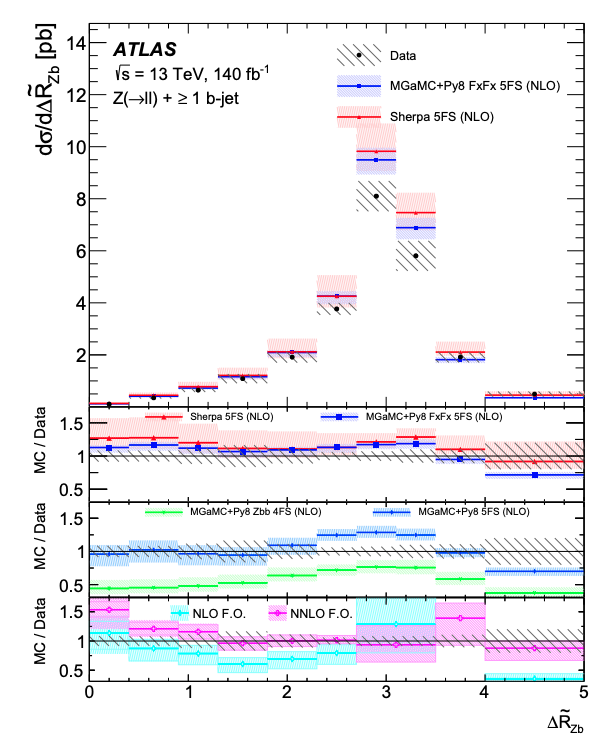}
    \caption{}
    \label{fig:Zb}
  \end{subfigure}
  \hfil
  \begin{subfigure}[b]{0.2\textwidth}
    \centering
    \includegraphics[scale=0.18]{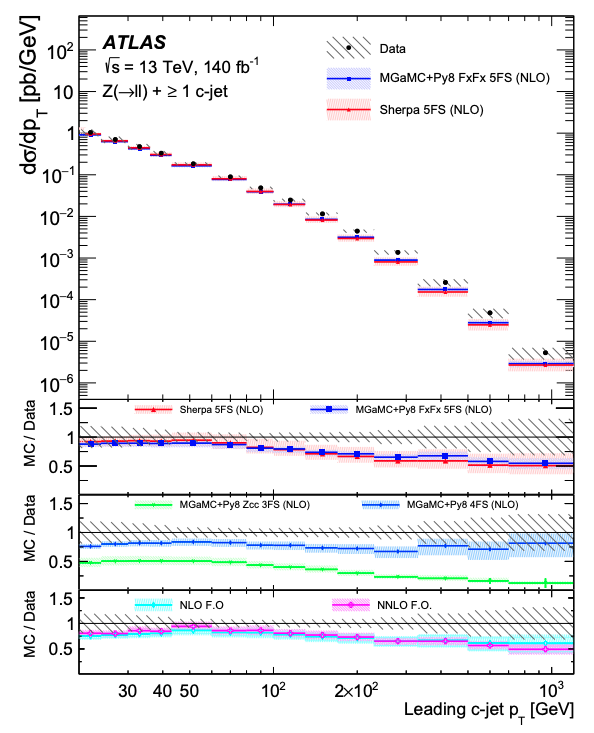}
    \caption{}
    \label{fig:Zc}
  \end{subfigure}
  \caption{(a) Measured fiducial cross section for $Z$+b-jets production as a function of $\Delta \tilde{R}_{Zb}$~\cite{ATLAS:Z+c}. \\
  (b) Measured fiducial cross section for $Z$+c-jets production as a function of the leading c-jet $p_T$~\cite{ATLAS:Z+c}.}
  \label{fig:Zbc}
\end{figure}

\noindent
The measurement of $Z$ boson production in association with jets originating from bottom or charm quarks in proton--proton collisions plays a key role in testing perturbative QCD (pQCD) and probing the internal structure of the proton. \\
$Z$+c- and $Z$+b-jet processes constitute relevant background contributions to several important analyses, such as Higgs boson measurements and searches for new physics. These measurements also provide crucial input for improving Monte Carlo (MC) event generator modelling. \\
Measurements of the inclusive and differential production cross sections of a $Z$ boson, decaying into electrons or muons, in association with at least one c-jets, at least one b-jet, or at least two b/c-jets, have been performed using the full Run~2 proton--proton dataset collected by the ATLAS experiment at $\sqrt{s}=13$~TeV~\cite{ATLAS:Z+c}. \\
Differential cross sections are measured for several observables sensitive to pQCD effects, PDF models, and MC generator validation. \\
In particular, the transverse momentum $p_T$ of the leading jet provides sensitivity to both pQCD predictions and MC modelling. \\
The $Z$+b-jet and $Z$+c-jet MC predictions are available in different flavour and mass schemes. 
The flavour-number schemes include the four-flavour (4FS) and five-flavour (5FS) approaches. 
In the 4FS, b-quarks are generated only through gluon splitting and are treated as massive; they do not contribute to the proton PDFs. 
In the 5FS scheme, the b-quark is included in the PDF and the approximation is more suitable for processes where the energy scale is much larger than the b-quark mass. \\
The observable $\Delta \tilde{R}_{Zb}$, defined as the angular distance between the $Z$ boson and the leading b-jet, is particularly sensitive to differences between flavour-scheme calculations. \\
In this analysis, events are selected containing a leptonically decaying $Z$ boson (into electrons or muons) in association with at least one or two jets containing a b- or c-hadron. 
Multijet and top-quark backgrounds are estimated using data-driven techniques, while other backgrounds are estimated using MC simulation. 
Selected events are categorized as $Z$+b-jets, $Z$+c-jets, or $Z$+light-jets through a likelihood fit to a flavour-sensitive observable. 
After background subtraction, the data are unfolded and compared with predictions from several MC generators, as well as fixed-order NNLO calculations. \\
Figure~\ref{fig:Zb} shows the measured fiducial cross section for $Z$+b-jet production as a function of $\Delta \tilde{R}_{Zb}$, compared with various flavour-scheme and (N)NLO predictions. 
The error bars represent the statistical uncertainty, while the hatched bands indicate the combined statistical and systematic uncertainties added in quadrature. 
The shaded bands correspond to the prediction uncertainties. 
In general, \texttt{Sherpa~2.2.11} and \texttt{MGaMC+PY8~FxFx} provide good agreement with the data, slightly overestimating it in the region corresponding to the $Z$+b back-to-back topology. 
\texttt{MGaMC+PY8~4FS~(NLO)} shows the poorest agreement: in the tail of the distribution, corresponding to collinear configurations and large $\Delta \tilde{R}_{Zb}$ values, the predicted topologies underestimate the data by more than 50\%. 
The NLO calculations underestimate the data for $\Delta \tilde{R}_{Zb}$ between 1 and 2.5 and above 4, while the agreement improves significantly at NNLO. \\
Figure~\ref{fig:Zc} shows the measured fiducial cross section for $Z$+c-jet production as a function of the leading c-jet $p_T$. 
The uncertainty conventions are the same as in Figure~\ref{fig:Zb}. 
The spectrum is well described by \texttt{Sherpa~2.2.11} and \texttt{MGaMC+PY8~FxFx} at low $p_T$, while at higher $p_T$ values the data are significantly underestimated. 
A better description is provided by \texttt{MGaMC+PY8~4FS~(NLO)}, although it remains close to or below the lower edge of the data uncertainty band. 
The \texttt{MGaMC+PY8~3FS~(NLO)} prediction consistently lies below the data. 
Both the NLO and NNLO fixed-order predictions yield spectra softer than observed in the data. \\

\section*{Conclusions}

\noindent
An overview of recent ATLAS measurements of multijet and vector-boson–plus–jets production has been presented. 
These studies provide a deeper understanding of Quantum Chromodynamics (QCD) and Electroweak (EW) interactions, testing theoretical predictions across a wide range of energy scales. \\
Two of the analyses discussed focus on jet substructure observables, which have become increasingly important for exploring the hadronization process and the internal structure of jets. 
The \textit{Measurement of jet track functions in $pp$ collisions at $\sqrt{s}=13$~TeV with the ATLAS detector}~\cite{ATLAS:2923297} and the \textit{Measurement of the Lund Jet Plane in hadronic decays of top quarks and $W$ bosons with the ATLAS detector}~\cite{ATLAS:LJP} provide new insights into parton shower evolution, non-perturbative effects, and the modeling of boosted topologies. \\
The other two analyses concern the production of vector bosons in association with multiple jets, which are sensitive to both QCD and EW higher-order corrections and offer valuable constraints on parton distribution functions (PDFs). 
The \textit{Cross-section measurements for the production of a $W$ boson in association with high-transverse-momentum jets in $pp$ collisions at $\sqrt{s}=13$~TeV with the ATLAS detector}~\cite{ATLAS:2920036} and the \textit{Measurements of the production cross-section for a $Z$ boson in association with $b$- or $c$-jets in proton–proton collisions at $\sqrt{s}=13$~TeV with the ATLAS detector}~\cite{ATLAS:Z+c} extend previous results and probe regions where QCD radiation, flavour schemes, and EW corrections play a significant role. \\
Overall, these measurements provide stringent tests of the current theoretical models and Monte Carlo event generators, highlighting areas where further tuning and theoretical improvements are required. 
They represent an important step toward a more precise description of multijet and vector-boson–plus–jets processes at the LHC, and lay the groundwork for future Run~3 and High-Luminosity LHC analyses.

%
%

\bibliographystyle{plain}

	\bibliography{jacow}
%
%

\end{document}